\documentstyle[12pt]{article}
\begin{document}
\tolerance=10000

\title{Intrinsic and extrinsic vortex nucleation mechanisms in the flow}
\author{F.V. Kusmartsev;
 NORDITA, Blegdamsvej 17, DK-2100 Copenhagen 0, Denmark;
 Department of  Physics,
   Loughboro' University
 Loughboro', Leicestershire, LE11 3TU, UK}
\maketitle

\begin{abstract}{%
We propose very general   vortex nucleation mechanisms\cite{KusVort}
analogous to a hydrodynamic instability and calculate 
associated critical velocity in agreement with experiments.
 The creation of vortices via extrinsic mechanism
is driven by a formation of the surface vorticity sheet 
created by the flow, which reaches a critical
size. Such a sheet screens an attraction of a half-vortex ring to the wall, the barrier for the vortex
nucleation disappears and the vortex nucleation
is started. 
In the intrinsic mechanism the creation of a big vortex ring, which transforms into the vortex,
is driven by a fluctuative generation of small vortex rings.}
\end{abstract}

\vskip 0.5 truecm

    Contributed paper to the  XXI International Conference on Low Temperature
Physics, August 8-14, 1996.

\vfill \eject

 Numerous experiments in superfluid $^3$He  and $^4$He show 
that critical velocity of the superflow is described
by the scaling expression  $V_c=V_0(1-T/T_c)^p$ with
the scaling exponent $p$\cite{Pack,Godn}. This universality
can not be described in the framework Iordanskii- Langer- Fisher
activational nucleation theory (see, also, references in\cite{Godn}).
To remove this discrepancy we propose new type 
of vortex nucleation mechanisms\cite{KusVort} relevant both
for superfluids and for superconductors which  
 may be both
{\it intrinsic and extrinsic}.
The mechanisms are  based
on the creation of the stochastic regions with fluctuative vorticity \cite{KusVort}. 
 The vortex penetrates the nucleation barrier with the aid of
 critical fluctuations via a creation of  the fluctuative
 vorticity regions.
The size of the regions depends on the {\it flow velocity}. 
 When the size of these regions
 reaches a critical value
a barrierless vortex nucleation is started. The critical size depends
only on the temperature.
  
  {\it The vortex nucleates} in a process similar to the
Berezinskii-Kosterlitz-Thouless (BKT) phase transition 
\cite{BKT}, where instead of
  vortex-antivortex (V-A) pairs  in the nucleation region  a fluctuative  vorticity liquid  is generated.
  A similar  situation  occurs in the Williams-Shenoy(WS)
\cite {Will} model of the $\lambda$-phase transition where the role of V-A pairs of the BKT transition is
played by  the  vortex rings.
In the same spirit taking into account the vorticity fluctuations on smaller scale
in the nucleation region, with the aid of renormalization group
we  obtain the 
 scaling
universality for the critical velocity  of the form \cite{KusVort}:
\begin{equation}
V_c=V_0(1-T/T_c)^p,
\label{crit-vel}
\end{equation}
where the value $V_0=\hbar/m a_c$ and $ a_c$ is a vortex core radius at zero
temperature. For $^4$He it is  $a_c=2.0-2.3$\AA  \cite{SaKu}. The critical exponent $p$ depends on the type of 
a vortex nucleation or on the dominating shape of the vorticity
fluctuations. We derived the renormalization group taking into account
 both  the vortex loops and  the half-loops fluctuations. 
The derivation of 
scaling relations is associated with the problem of
 two relevant operators:
 the temperature and
the flow velocity. This problem is solved due to a finite size
of the critical fluctuation regions. That is the scaling 
associated with the flow velocity and temperature
is stopped when the size of 
the nucleation center reaches the critical value, at which the
barrier for the vortex nucleation disappears.
The nucleation centers are generated
by the flow and
characterize {\it  a hydrodynamic threshold instability}.

{\it The extrinsic mechanism}  is related
to some centers of nucleation around which a surface vorticity
sheet is created. Such centers of nucleations may be both
extended and  pointlike. In the first case it is a smooth
surface of the rotating bucket or smooth walls 
in small apertures, in orifices or in narrow channels
of Vycor glasses. The pointlike centers of nucleation
may be a surface asperity on smooth walls or impurity atoms
like $^3$He impurities in $^4$He.

For {\it the superflow  near a smooth surface} there occurs
a hydrodynamic instability analogous to a surface phase transition, in which 
{\it the  width of  the surface vorticity sheet 
 reaches a critical size $R_c$},
the barrier for the vortex nucleation  disappears and the vortex generation
 is started.  As the flow velocity
increases the energy of the pinned  half-vortex ring  decreases.
This stimulates
their activation through thermal fluctuations.
 In their turn the
fluctuational half-vortex rings  of a small radius  assist
in the creation of
 half-vortex rings  of larger radius and so on.
The picture is reminiscent of
the scaling in BKT transition, where 
 the coupling between the vortex and
antivortex  decreases as the temperature and the flow rise.
 The scaling relation for critical velocity  again takes the derived
universal form (\ref{crit-vel}) with $p=1$.

{\it In  Vycor glasses}, where one has narrow
channels instead of
the half-vortex rings for the plane geometry,
  the optimal shape of fluctuations
 will be  small segments
 of the vortex rings. This shape depends on the curvature, i.e. on the
radius of the narrow channel. Because of this  the critical temperature of the phase transition  decreases while the critical indices   i.e.  the universality class of the phase transition
 remains the same and  the critical velocity
 is described by the same critical exponent with $p=1$.

If there is {\it a microscopic surface asperity}
 (a small mountain peak) on a smooth surface it may play the most important
role in the  nucleation of vortices, since in its neighborhood
 the flow is the greatest. On this site  the surface vorticity cloud
is created more efficiently and it has bigger radius than the surface
vorticity sheet on a smooth surface. This surface vorticity cloud 
screens locally ( in the region of the surface asperity)
 the attraction of the half-vortex ring to the wall. 
When the size of this vorticity cloud reaches the critical value
 the vortex barrierless nucleation  is started.
This is described by the same universal 
scaling relation, eq.(1) with $p=1$ \cite{KusVort,Pack}.
The scaling exponent 
has not been changed due to the universal character
of the creation of critical stochastic regions.
 In this case the role of aperture
size is played by the size of the surface asperity.
 In the rotating bucket experiments  the nucleation mechanism is identical to the aperture experiments.

 {\it The role of the impurity atoms} (like $^3$He  in the $^4$He) or 
 clusters of impurity atoms (if any exist) in the vortex 
nucleation is similar to a surface asperity, i.e. they serve as  centers of
the stochastic regions filled with the surface vorticity. 
That is any inhomogeneity in the bulk similar to the
 surface asperity stimulates locally the vorticity cloud and serves as a
 vortex nucleation center. Such situations occur only if
 there are very rare density of
  surface asperities or
a very low density of surface inhomogeneities.
If the density of such inhomogeneities increases and will create some
random potential, then the correlations of this random potential
may, in principle, change the value of the scaling exponent $p$.
  This may be relevant for Aerogel and Xerogel glasses \cite{Pack}.

 {\it In the intrinsic mechanism}
 the vortices are created in a flow of superfluid
with the aid of nucleation of droplets filled with a fluctuative
vorticity liquid. The optimal dominating
shape of vorticity fluctuations  is a vortex ring.
Taking into account such vorticity fluctuations we got the critical
exponent $p=1$, which  is in a poor agreement with experiments in
$^4$He \cite{Godn}. If one considers the bending of the fluctuative
vortex rings in the droplets  
 one must use Flory arguments
about self-avoiding walk exponent, i.e., to take into account 
 the vortex tangle or
 bending of the loops. This may change the value of $p$.

All these arguments with scaling do  work in $^3$He: where, however,
 the thermal fluctuation region is very close to $T_c$. But since the
 vortices are created due to hydrodynamic 
flow instability\cite{KusVort}, the critical exponent $p=1/2$,
(the surface roughness changes this to $p=1/4$ \cite{Godn}) is valid not too close to $T_c$.

I thank  P.V.E. McClintock, R.E. Packard and  G.E. Volovik for a fruitful  discussion.
The work supported by 
NORDITA.

\ \\

also address: Landau Institute for Theoretical Physics, Moscow, Russia

\vfill
\eject

\end{document}